\documentclass[a4paper,11pt]{article}
\usepackage[T1]{fontenc} 
\usepackage{float} 
\usepackage{subfigure}
\usepackage{ragged2e}
\usepackage{amsmath}
\usepackage{amssymb}
\usepackage{array}
\usepackage{makecell}
\usepackage{color}
\usepackage{diagbox}
\usepackage{multirow}
\makeatletter
\gdef\@fpheader{ }
\gdef\@journal{ }
\makeatother
\RequirePackage{amsmath}
\RequirePackage{amssymb}
\RequirePackage{epsfig}
\RequirePackage{graphicx}
\RequirePackage[numbers,sort&compress]{natbib}
\RequirePackage{color}
\RequirePackage[colorlinks=true
,urlcolor=blue
,anchorcolor=blue
,citecolor=blue
,filecolor=blue
,linkcolor=blue
,menucolor=blue
,pagecolor=blue
,linktocpage=true
,pdfproducer=medialab
,pdfa=true
]{hyperref}

\newif\ifnotoc\notocfalse
\newif\ifemailadd\emailaddfalse
\newif\iftoccontinuous\toccontinuousfalse
\makeatletter
\def\@subheader{\@empty}
\def\@keywords{\@empty}
\def\@abstract{\@empty}
\def\@xtum{\@empty}
\def\@dedicated{\@empty}
\def\@arxivnumber{\@empty}
\def\@collaboration{\@empty}
\def\@collaborationImg{\@empty}
\def\@proceeding{\@empty}
\def\@preprint{\@empty}

\newcommand{\subheader}[1]{\gdef\@subheader{#1}}
\newcommand{\keywords}[1]{\if!\@keywords!\gdef\@keywords{#1}\else%
\PackageWarningNoLine{\jname}{Keywords already defined.\MessageBreak Ignoring last definition.}\fi}
\renewcommand{\abstract}[1]{\gdef\@abstract{#1}}
\newcommand{\dedicated}[1]{\gdef\@dedicated{#1}}
\newcommand{\arxivnumber}[1]{\gdef\@arxivnumber{#1}}
\newcommand{\proceeding}[1]{\gdef\@proceeding{#1}}
\newcommand{\xtumfont}[1]{\textsc{#1}}
\newcommand{\correctionref}[3]{\gdef\@xtum{\xtumfont{#1} \href{#2}{#3}}}
\newcommand\jname{JHEP}

\newcommand\preprint[1]{\gdef\@preprint{\hfill #1}}

\makeatother

\newcommand\note[2][]{%
\if!#1!%
\stepcounter{footnote}\footnotetext{#2}%
\else%
{\renewcommand\thefootnote{#1}%
\footnotetext{#2}}%
\fi}

\makeatletter
\newtoks\auth@toks
\renewcommand{\author}[2][]{%
  \if!#1!%
    \auth@toks=\expandafter{\the\auth@toks#2\ }%
  \else
    \auth@toks=\expandafter{\the\auth@toks#2$^{#1}$\ }%
  \fi
}
\makeatother
\makeatletter
\newtoks\affil@toks\newif\ifaffil\affilfalse
\newcommand{\affiliation}[2][]{%
\affiltrue
  \if!#1!%
    \affil@toks=\expandafter{\the\affil@toks{\item[]#2}}%
  \else
    \affil@toks=\expandafter{\the\affil@toks{\item[$^{#1}$]#2}}%
  \fi
}
\makeatother
\makeatletter
\newtoks\email@toks\newcounter{email@counter}%
\newcommand{\emailAdd}[1]{%
\emailaddtrue%
\ifnum\theemail@counter>0\email@toks=\expandafter{\the\email@toks, \@email{#1}}%
\else\email@toks=\expandafter{\the\email@toks\@email{#1}}%
\fi\stepcounter{email@counter}}
\newcommand{\@email}[1]{\href{mailto:#1}{\tt #1}}
\makeatother

\makeatletter
\newcommand*\collaboration[1]{\gdef\@collaboration{#1}}
\newcommand*\collaborationImg[2][]{\gdef\@collaborationImg{#2}}
\makeatletter
\newcommand\afterLogoSpace{\smallskip}
\newcommand\afterSubheaderSpace{\vskip3pt plus 2pt minus 1pt}
\newcommand\afterProceedingsSpace{\vskip21pt plus0.4fil minus15pt}
\newcommand\afterTitleSpace{\vskip23pt plus0.06fil minus13pt}
\newcommand\afterRuleSpace{\vskip23pt plus0.06fil minus13pt}
\newcommand\afterCollaborationSpace{\vskip3pt plus 2pt minus 1pt}
\newcommand\afterCollaborationImgSpace{\vskip3pt plus 2pt minus 1pt}
\newcommand\afterAuthorSpace{\vskip5pt plus4pt minus4pt}
\newcommand\afterAffiliationSpace{\vskip3pt plus3pt}
\newcommand\afterEmailSpace{\vskip16pt plus9pt minus10pt\filbreak}
\newcommand\afterXtumSpace{\par\bigskip}
\newcommand\afterAbstractSpace{\vskip16pt plus9pt minus13pt}
\newcommand\afterKeywordsSpace{\vskip16pt plus9pt minus13pt}
\newcommand\afterArxivSpace{\vskip3pt plus0.01fil minus10pt}
\newcommand\afterDedicatedSpace{\vskip0pt plus0.01fil}
\newcommand\afterTocSpace{\bigskip\medskip}
\newcommand\afterTocRuleSpace{\bigskip\bigskip}
\newlength{\affiliationsSep}
\newcommand\beforetochook{\pagestyle{myplain}\pagenumbering{roman}}

\DeclareFixedFont\trfont{OT1}{phv}{b}{sc}{11}

\renewcommand\maketitle{
\pagestyle{empty}
\thispagestyle{titlepage}
\setcounter{page}{0}
\noindent{\small\scshape\@fpheader}\@preprint\par

\afterLogoSpace
\if!\@subheader!\else\noindent{\trfont{\@subheader}}\fi
\afterSubheaderSpace
\if!\@proceeding!\else\noindent{\sc\@proceeding}\fi
\afterProceedingsSpace
{\LARGE\flushleft\sffamily\bfseries\@title\par}
\afterTitleSpace
\hrule height 1.5\p@%
\afterRuleSpace
\if!\@collaboration!\else
{\Large\bfseries\sffamily\raggedright\@collaboration}\par
\afterCollaborationSpace
\fi
\if!\@collaborationImg!\else
{\normalsize\bfseries\sffamily\raggedright\@collaborationImg}\par
\afterCollaborationImgSpace
\fi
{\bfseries\raggedright\sffamily\the\auth@toks\par}
\afterAuthorSpace
\ifaffil\begin{list}{}{%
\setlength{\leftmargin}{0.28cm}%
\setlength{\labelsep}{0pt}%
\setlength{\itemsep}{\affiliationsSep}%
\setlength{\topsep}{-\parskip}}
\itshape\small%
\the\affil@toks
\end{list}\fi
\afterAffiliationSpace
\ifemailadd 
\noindent\hspace{0.28cm}\begin{minipage}[l]{.9\textwidth}
\begin{flushleft}
\textit{E-mail:} \the\email@toks
\end{flushleft}
\end{minipage}
\else 
\PackageWarningNoLine{\jname}{E-mails are missing.\MessageBreak Plese use \protect\emailAdd\space macro to provide e-mails.}
\fi
\afterEmailSpace
\if!\@xtum!\else\noindent{\@xtum}\afterXtumSpace\fi
\if!\@abstract!\else\noindent{\renewcommand\baselinestretch{.9}\textsc{Abstract:}}\ \@abstract\afterAbstractSpace\fi
\if!\@keywords!\else\noindent{\textsc{Keywords:}} \@keywords\afterKeywordsSpace\fi
\if!\@arxivnumber!\else\noindent{\textsc{ArXiv ePrint:}} \href{http://arxiv.org/abs/\@arxivnumber}{\@arxivnumber}\afterArxivSpace\fi
\if!\@dedicated!\else\vbox{\small\it\raggedleft\@dedicated}\afterDedicatedSpace\fi
\ifnotoc\else
\iftoccontinuous\else\newpage\fi
\beforetochook\hrule
\tableofcontents
\afterTocSpace
\hrule
\afterTocRuleSpace
\fi
\setcounter{footnote}{0}
\pagestyle{myplain}\pagenumbering{arabic}
} 

\renewcommand{\baselinestretch}{1.1}\normalsize
\divide\@tempdima\baselineskip
\@tempcnta=\@tempdima
\skip\@mpfootins = \skip\footins
\renewcommand{\@dotsep}{10000}

\newcommand\ps@myplain{
\pagenumbering{arabic}
\renewcommand\@oddfoot{\hfill-- \thepage\ --\hfill}
\renewcommand\@oddhead{}}
\let\ps@plain=\ps@myplain

\newcommand\ps@titlepage{\renewcommand\@oddfoot{}\renewcommand\@oddhead{}}

\numberwithin{equation}{section}

\renewcommand\section{\@startsection{section}{1}{\z@}%
                                   {-3.5ex \@plus -1.3ex \@minus -.7ex}%
                                   {2.3ex \@plus.4ex \@minus .4ex}%
                                   {\normalfont\large\bfseries}}
\renewcommand\subsection{\@startsection{subsection}{2}{\z@}%
                                   {-2.3ex\@plus -1ex \@minus -.5ex}%
                                   {1.2ex \@plus .3ex \@minus .3ex}%
                                   {\normalfont\normalsize\bfseries}}
\renewcommand\subsubsection{\@startsection{subsubsection}{3}{\z@}%
                                   {-2.3ex\@plus -1ex \@minus -.5ex}%
                                   {1ex \@plus .2ex \@minus .2ex}%
                                   {\normalfont\normalsize\bfseries}}
\renewcommand\paragraph{\@startsection{paragraph}{4}{\z@}%
                                   {1.75ex \@plus1ex \@minus.2ex}%
                                   {-1em}%
                                   {\normalfont\normalsize\bfseries}}
\renewcommand\subparagraph{\@startsection{subparagraph}{5}{\parindent}%
                                   {1.75ex \@plus1ex \@minus .2ex}%
                                   {-1em}%
                                   {\normalfont\normalsize\bfseries}}

\def\fnum@figure{\textbf{\figurename\nobreakspace\thefigure}}
\def\fnum@table{\textbf{\tablename\nobreakspace\thetable}}

\long\def\@makecaption#1#2{%
  \vskip\abovecaptionskip
  \sbox\@tempboxa{\small #1. #2}%
  \ifdim \wd\@tempboxa >\hsize
    \small #1. #2\par
  \else
    \global \@minipagefalse
    \hb@xt@\hsize{\hfil\box\@tempboxa\hfil}%
  \fi
  \vskip\belowcaptionskip}

\renewenvironment{thebibliography}[1]{%
\begin{oldthebibliography}{#1}%
\small%
\raggedright%
\setlength{\itemsep}{5pt plus 0.2ex minus 0.05ex}%
}%
{%
\end{oldthebibliography}%
}

\begin{document}


\title{\boldmath A Data Augmentation Method and the Embedding Mechanism 
for Detection and Classification of Pulmonary Nodules on Small Samples}

\author[a,b,1]{Yang Liu,}\note{Yue-Jie Hou, and Yang Liu contributed equally to this work.}
\author[a,1]{Yue-Jie Hou,}
\author[a]{Chen-Xin Qin,}
\author[a]{Xin-Hui Li,}
\author[a]{Qi-Meng Du,}
\author[a,2]{Si-Jing Li,}
\author[c,d,e,2]{Bin Wang,}
\author[a,2]{Chi-Chun
Zhou}\note{Corresponding author. sijing\_li@dali.edu.cn; wangbin23@csu.edu.cn; zhouchichun@dali.edu.cn}

\affiliation[a]{School of Engineering, Dali University, Dali, Yunnan 671003, PR China;}
\affiliation[b]{School of Electronic and Information Engineering, Liuzhou Vocational \& Technical College, Liuzhou China, PR China;}
\affiliation[c]{Department of Thoracic Surgery, The Second Xiangya Hospital of Central
South University, Changsha, China; No. 139 Renmin Road,Changsha, Hunan, PR China;}
\affiliation[d]{ Hunan Key Laboratory of Early Diagnosis and Precise Treatment of Lung Cancer, The Second Xiangya Hospital of Central South University, Changsha,
China; No. 139 Renmin Road,Changsha, Hunan, PR China; }
\affiliation[e]{Early-Stage Lung Cancer Center, The Second Xiangya Hospital of Central
South University, Changsha, China; No. 139 Renmin Road,Changsha, Hunan, PR China;}











\abstract{Introduction:Detection of pulmonary nodules by CT is used for screening lung cancer in early stages.omputer aided diagnosis (CAD) based on deep-learning method can identify the suspected areas of pulmonary nodules in CT images, thus improving the accuracy and efficiency of CT diagnosis. The accuracy and robustness of deep learning models. 
Method:In this paper, we explore (1) the data augmentation method based on the generation model and (2) the model structure improvement method based on the embedding mechanism. Two strategies have been introduced in this study: a new data augmentation method and a embedding mechanism. In the augmentation method, a 3D pixel-level statistics algorithm is proposed to generate pulmonary nodule and by combing the faked pulmonary nodule and healthy lung, we generate new pulmonary nodule samples. The embedding mechanism are designed to better understand the meaning of pixels of the pulmonary nodule samples by introducing hidden variables. 
Result: The result of the 3DVNET model with the augmentation method for pulmonary nodule detection shows that the proposed data augmentation method outperforms the method based on generative adversarial network (GAN) framework, training accuracy improved by 1.5$\%$, and with embedding mechanism for pulmonary nodules classification shows that the embedding mechanism improves the accuracy and robustness for the classification of pulmonary nodules obviously, the model training accuracy is close to 1 and the model testing F1-score is 0.90.
Conclusion:he proposed data augmentation method and embedding mechanism are beneficial to improve the accuracy and robustness of the model, and can be further applied in other common diagnostic imaging tasks. }


\maketitle
\flushbottom


\section{Introduction}
Lung cancers have the highest mortality among various cancers with high incidence rate
\cite{lutong2018,gaochufan2019}.
Pulmonary nodules are the initial manifestation of the vast majority of lung cancer in radiologic diagnosis. 
Due to Computed Tomography (CT), which has high
resolution and is fast and easy to use, doctors can use a patient’s CT data to learn about the grays, textures, and densities of the body’s tissues and to detect subtle differences. And
then predict whether there are early lesions - lung nodules.

Computer-Aided Diagnosis (CAD)\cite{yanase2019}, which is based on computer-aided medical diagnosis, has been used to help doctors analyze CT data. After each patient’s CT scan
is completed, there will be hundreds of images with a lot of information, which creates
a burden for doctors. CAD system can distinguish suspected areas of pulmonary nodules
more efficiently and accurately, and is used to assist doctors in diagnosis. 
The deep learning models for detection and classification of pulmonary nodules
can improve productivity.
The traditional methods of pulmonary nodule 
detection depend on the manual extraction of pulmonary nodule features and the selection of 
useful feature training \cite{lynch2018}.
With the wide application of deep 
learning and its combination with the field of image processing,
deep learning algorithm has gradually become a hot research topic.
Deep learning algorithm can automatically extracts image feature more accurately and comprehensively based on deep network \cite{yu2015}.
At present, the diagnostic methods of pulmonary nodules mainly combine traditional
machine learning algorithm with deep learning, such as the application
of traditional algorithm in the segmentation of suspected pulmonary
nodules region, and use a deep convolutional neural network in
the removal of false positive pulmonary nodules\cite{ohno2020}.

In the medical field,
in order to further improve the accuracy and robustness of deep learning models 
for detection and classification of pulmonary nodules,
it can be achieved by increasing the data and improving the model structure.
However, the lack of medical image data makes it difficult to apply deep learning methods in this field.
In terms of increasing data,
the traditional method of data augmentation is usually physical transformation of image data,
such as rotation, cutting and flipping \cite{setio2016},
traditional approach is essentially in repeated training samples of the existing features.
At present,
the data augmentation method based on generation model is widely studied \cite{guimaraes2017}.
Different from traditional data augmentation methods,
the generative model learn the distribution of features in the training samples,
and then generation new samples increase the amount of data,
so can effectively improve accuracy and robustness of the model.
The main methods of data augmentation based on generation model include:
building the model of generating target data with hidden variables \cite{guimaraes2017},
generative adversarial network (GAN) framework \cite{goodfellow2014},
and changing GANs \cite{yixin2019,tanjiaxing2021},
etc.
To improve the structure of the model,
people expand the traditional 2-dimensional convolution to 3-dimensional \cite{kang2017} with
deeper network structure and integrate various mechanisms,
such as the transfer learning \cite{ravishankar2016} combined with the priori information,
residual mechanism \cite{hekaiming2016}, multi-model mechanism \cite{xiaoyawen2018},
multi-task learning approach \cite{chen2019}, etc.
All of these methods have been proved usefully for improving the performance of the model within a certain range.
Many medical breakthroughs have occurred with self-experimentation and single experiments. Studies, particularly analytical studies, may provide more truthful results with a small sample because intensive efforts can be made to control all the confounders, wherever they operate, and sophisticated equipment can be used to obtain more accurate data \cite{Indrayan2021}.No single study, whether based on a small sample or a large sample, is considered conclusive. A large number of small studies can be done easily in different setups, and if they point toward the same direction, a safe, possibly more robust, conclusion can be drawn through a meta-analysis\cite{Anderson2001, Wasserstein2019}.

In this work, we use small samples to explore the data augmentation method based 
on generated model and the model structure improvement method 
based on hidden variable embedding mechanism. 

\begin{figure}[H]
	\centering
	\includegraphics[width=1\textwidth]{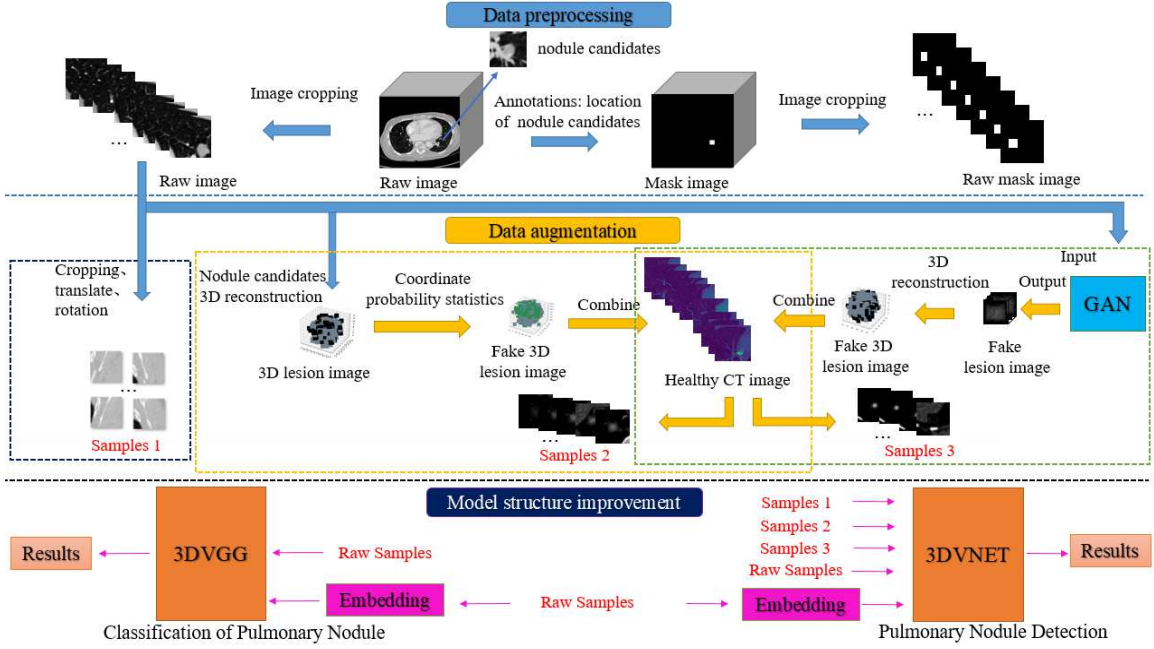}
	\caption{flow of the paper.}
	\label{all}
\end{figure}

 The work is organized as follows: in Sec.2, we introduce the datasets,including the original dataset and the dataset obtained by using three data augmentation methods. in Sec.3, we introduce the framework of pulmonary nodules detection and classification, and the technique of pixel embedding. in Sec.4, we test the effect of introducing the embedding mechanism based on hidden variables into the pulmonary nodule classification and detection model on the experimental result,and test the effect of lung nodule samples generated using two data augmentation methods on the results of the detection model. 
and outlooks are given in Sec.5 .
The flow of the paper is shown in Figure~\ref{all}.

\section{Dataset}

\subsection{LUNA16 Data}

 

The LIDC-IDRI \cite{jacobs2016} database collected by the National Cancer Institute of the United States.
 A total of $1010$ patient samples were included in the dataset,
 including $243958$ images.
 The database annotates the CT image file where the nodules are located and the specific coordinates.
 Researchers can use annotated documents to research on the detection of pulmonary nodules,
 the LIDC-IDRI database is a commonly used database for pulmonary nodules detection.

The LUNA16 database \cite{setio2017} removed CT slices
 with section thickness greater than $ 3mm $ from the LIDC-IDR database \cite{heelan1984},
 and removed the inconsistent section step size and missing ones from the CT images.
 The Luna$16$ database has 888 cases,
 and 888 cases of CT data were evenly divided into 10 folders from subset$0$ to subset$9$.
 A CT image is composed of multiple two-dimensional slices,
 which form a three-dimensional data (Z,X,Y three different dimensions).
 CT information labeling was completed jointly by $4$ physicians.
 The annotation file annotation.csv
 contains all $1186$ nodule labeling information (including nodule size coordinates and diameter),
 and the annotation file candidate.csv contains $551065$ candidate nodule labeling information
 (including nodule size coordinates and category labels).
 \subsection{Small Sample Datasets}
In this section, we introduce the datasets structure of the study.     
In the medical field, the use of small samples for research is a common problem, many diseases with few positive cases,
so how to use small samples to train a good model becomes important.
The training pulmonary nodule detection model was selected from 13 cases of LUNA16,
1500 positive samples and 1500 negative samples are selected to generate 500 samples generated by adversarial network model and 500 samples generated by lung nodule generation algorithm based on 3D pixel-level statistics. 
In the testset, 649 positive samples and 7073 negative samples are selected from 4 cases in LUNA16. 
In order to balance the dataset, 50 positive samples and 50 negative samples are randomly selected for the experiment, and all the above samples were used to explore the influence of the data enhancement method on the model effect. 
In the evaluation experiment of pulmonary nodule classification model based on hidden variable embedding mechanism, 24042 negative samples and 50 positive samples from 5 cases of LUNA16 dataset are selected in the training set, and the 50 positive samples were processed by traditional data enhancement methods such as flip and rotation. 
In the final experiment, the positive sample is 12500, and the negative sample is 240,42. 
In the model testing stage, 2982 negative samples and 18 positive samples from 4 cases in LUNA16 dataset are selected, and the 18 positive samples are processed by traditional data augmentation methods such as flipping and rotation. 
In order to balance the dataset, 50 positive samples and 50 negative samples are randomly selected for the experiment to complete the evaluation experiment of the classification model. 
The detailed data set Settings are shown in Tables \ref{data31} to \ref{data34} .
\begin{table}[H]
\centering
\caption{Experimental training dataset for model evaluation of pulmonary nodule detection methods based on image augmentation and embedding mechanism.}
\centering
	\begin{tabular}{ccc}     
\hline
$\makecell[c]{The\, number\, of\, cases}$&$\makecell[c]{Number\,of\, positive\, samples\\in\,the\\experiment}$&$\makecell[c]{Number\,of \,negative \,samples\\in\,the\,experiment}$\\
\hline
$\makecell[c]{8}$&$ \makecell[c]{1000}$&$ \makecell[c]{1000}$\\
\hline
$\makecell[c]{5}$&$ \makecell[c]{500}$&$ \makecell[c]{500}$\\
\hline
\end{tabular}
\label{data31}
\end{table}

\begin{table}[H]
\centering
\caption{Experimental testing dataset for model evaluation of pulmonary nodule detection methods based on image augmentation and embedding mechanism.}
\centering
	\begin{tabular}{ccc}     
\hline
$\makecell[c]{The\, number\, of\, cases}$&$\makecell[c]{Number\,of\, positive\, samples\\in\,the\\experiment}$&$\makecell[c]{Number\,of \,negative \,samples\\in\,the\,experiment}$\\
\hline
$\makecell[c]{4}$&$ \makecell[c]{50}$&$ \makecell[c]{50}$\\

\hline
\end{tabular}
\label{data32}
\end{table}

\begin{table}[H]
\centering
\caption{The embedding mechanism based on hidden variables is used to evaluate the experimental training dataset of the pulmonary nodule classification model}
\centering
	\begin{tabular}{ccccc}     
\hline
$\makecell[c]{The\, number\, of\\ cases}$&$\makecell[c]{Total\,number\,of\\ positive\\ samples}$&$\makecell[c]{Total\,negative\\ number\,of \\samples}$&$\makecell[c]{The\,amount\,of\\ data\, after \\augmenting\,the\\ positive \\samples}$&$\makecell[c]{Number\,of \\negative \\samples\,in\,the\\experiment}$\\
\hline
$\makecell[c]{5}$&$ \makecell[c]{50}$&$ \makecell[c]{24042}$&$ \makecell[c]{12500}$&$ \makecell[c]{24042}$\\
\hline
\end{tabular}
\label{data33}
\end{table}

\begin{table}[H]
\centering
\caption{The embedding mechanism based on hidden variables is used to evaluate the experimental testing dataset of the pulmonary nodule classification model.}
\centering
	\begin{tabular}{ccccc}     
\hline
$\makecell[c]{The\, number\, of\\ cases}$&$\makecell[c]{Total\,number\,of\\ positive\\ samples}$&$\makecell[c]{Total\,negative\\ number\,of \\samples}$&$\makecell[c]{The\,amount\,of\\ data\, after \\augmenting\,the\\ positive \\samples}$&$\makecell[c]{Number\,of \\negative \\samples\,in\,the\\experiment}$\\
\hline
$\makecell[c]{4}$&$ \makecell[c]{18}$&$ \makecell[c]{2980}$&$ \makecell[c]{50}$&$ \makecell[c]{50}$\\
\hline
\end{tabular}
\label{data34}
\end{table}

\subsection{Data Augmentation Based on 3D Pixel-level Statistics Algorithm} 

\subsubsection{Three-dimensional Reconstruction Technique}
 
\begin{figure}[H]
	\centering
	\includegraphics[width=1\textwidth]{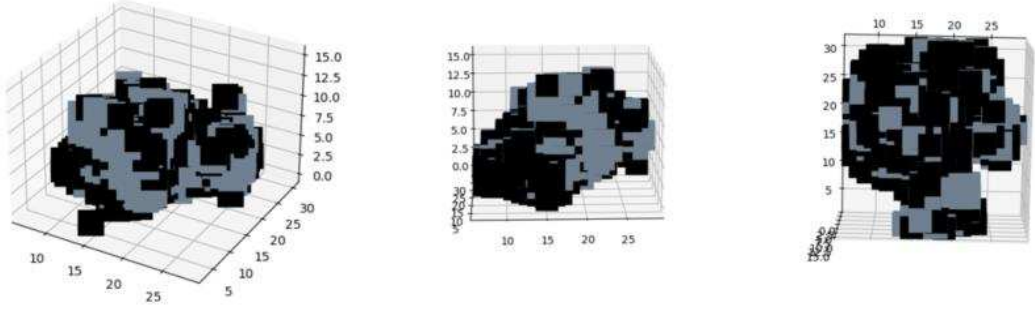}
	\caption{3D volume at 3 angles (3D nodules are separated from the background by threshold segmentation.}
	\label{jiejie}
\end{figure}

In this section,
a three-dimensional reconstruction technique is used to visualize the lung CT section and  nodules.
Three-dimensional reconstruction of medical data can make it
easier to understand the data information obtained in the experiment.
Images containing only nodules were obtained from CT images of Luna16 database.
In order to separate the nodule from the surrounding tissue,
the nodule coordinate region selected in the box was used to
find the center point with the same gray value according to the algorithm
for the center alignment of the two-dimensional image,
and then converted into the three-dimensional matrix through
the conversion function to establish the corresponding relationship
between the gray value of the nodule and the surrounding tissue \cite{lifrancis2017}.
The original image sequence is converted into a three-dimensional matrix sequence
to form a three-dimensional space containing gray values.
Finally, the 3D data of the nodule is reconstructed into 3D stereogram,
and the corresponding relationship of gray level is fine-tuned according to the 3D stereogram of the nodule,
so as to obtain a more accurate 3D stereogram of the pulmonary nodule. Figure~\ref{jiejie}
shows a three-dimensional view of the nodule.

\subsubsection{Prediction of 3D Pulmonary Nodules}
\begin{figure}[H]
	\centering
	\includegraphics[width=1\textwidth]{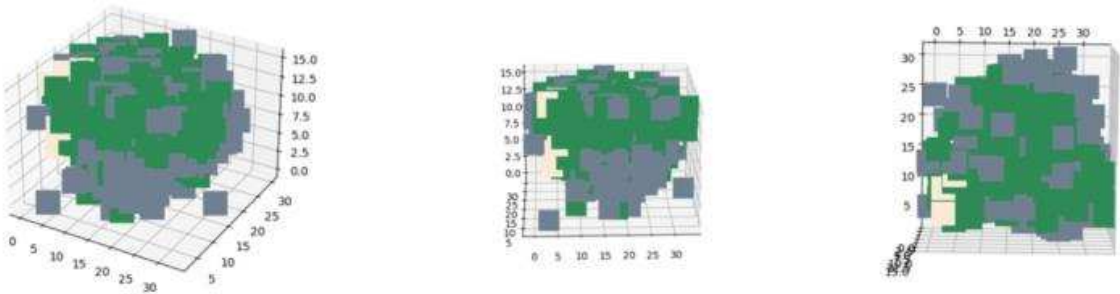}
	\caption{Predicts the generated three-dimensional nodule stereogram.}
	\label{glsc}
\end{figure}

The specific implementation of the lung nodule generation algorithm is as follows:
The image of the dataset is transformed into a 3D matrix containing gray value by 3D reconstruction.
Then, the three-dimensional data of multiple nodules were counted,
that is, the grayscale value of each three-dimensional coordinate point
and its occurrence times were converted into probability.
A sequence of two-dimensional images of pulmonary nodules is transformed into a three-dimensional matrix,
with each coordinate point $ (x,y,z) $ corresponding to a gray value.
The gray-scale values of the three-dimensional data of multiple different pulmonary nodules,
appearing in the same coordinate points and their occurrence times n were counted,
and the centers of them were all set to the same coordinate $ (x_{0},y_{0},z_{0}) $ for axis alignment.
The single coordinate $ (x,y,z) $ ,
the gray value of occurrence and the corresponding times are converted into probability,
as shown in Formula (\ref{gailv}).
The probability of the occurrence of the gray value in the corresponding coordinates
is used to generate the 3D matrix.
\begin{equation}
	P_{v_{i}(x,y,z)}=\dfrac{n_{i}}{sum{(n_{i})}} 
	\label{gailv}
\end{equation}
The gray values of all the 3D coordinate points obtained
are combined to form a 3D data for the prediction of nodules.
The three-dimensional nodule stereogram generated by the prediction is shown in Figure~\ref{glsc}.

\subsection{Data Augmentation Based on Generative Adversarial Networks}
\begin{figure}[H]
	\centering
	\includegraphics[width=1\textwidth]{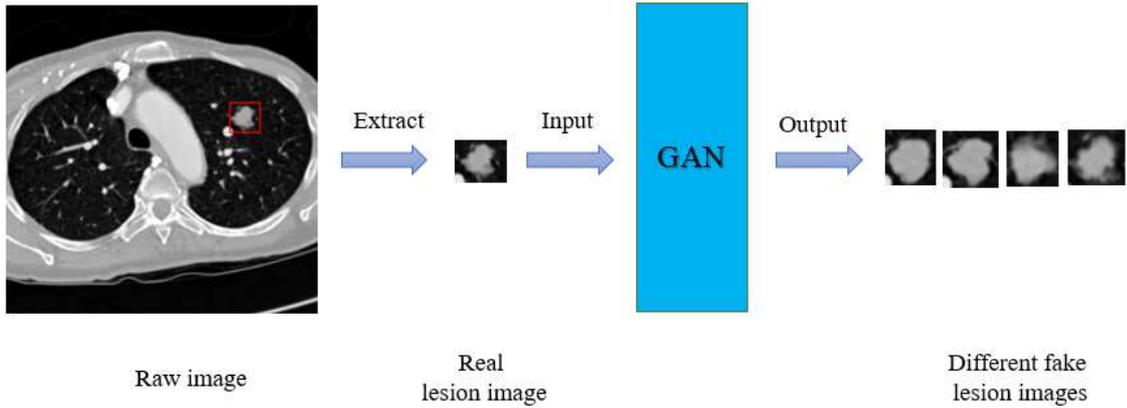}
	\caption{Flow chart of the method of data augmentation based on GAN.}
	\label{ganlc}
\end{figure}

Generative adversarial network (GAN) framework \cite{goodfellow2014} is a commonly used generation model,
it has the advantage of simulating the distribution of data,
In this section,
GAN is used to generate lung nodule data to increase the diversity of samples,
and this method can directly evaluate the quality of generated images.
The process is shown in Figure~\ref{ganlc}.

\begin{figure}[H]
	\centering
	\includegraphics[width=1\textwidth]{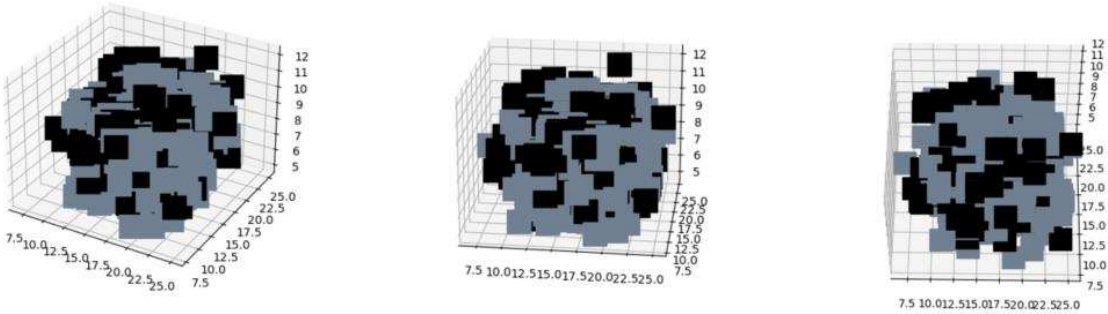}
	\caption{3D stereogram of nodules generated by GAN.}
	\label{gansw}
\end{figure}

In this section, we introduce the generated pulmonary nodules are reconstructed in three dimensions, and pulmonary nodule samples obtained by fusion of three-dimensional data pixels of pulmonary nodules with normal lung CT images.
Figure~\ref{gansw} shows a three-dimensional view of different pulmonary nodules generated by GAN framework.

\subsection{Pulmonary Nodule Smples Augmentation}
\begin{figure}[H]
	\centering
	\includegraphics[width=1\textwidth]{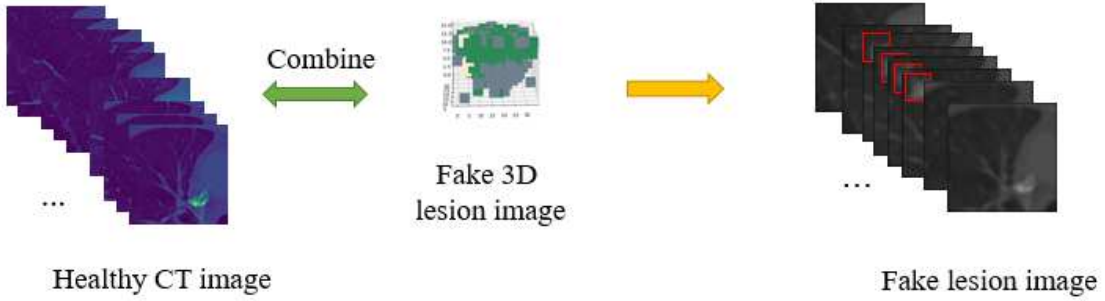}
	\caption{Flowchart of nodules embedded in healthy lung sections.}
	\label{jjqr}
\end{figure}

A new data augmentation method for generating pulmonary nodules was proposed.
The generated 3D pulmonary nodules are fused with normal lung CT images to obtain the new pulmonary nodule samples,
and the process shown in Figure~\ref{jjqr}.
The coordinate of a point generating pulmonary nodules is $ (x_{1},y_{1},z_{1}) $
and the gray value corresponding to the coordinate of this point is $ v_{1} $ .
Then, the gray value $ v_{2} $ corresponding to the same coordinate $ (x_{1},y_{1},z_{1}) $
in the lung section without nodules is replaced with $ v_{1} $ ,
and the gray value corresponding to all coordinate points
in the three-dimensional data generating pulmonary nodules is replaced by the same method.
All coordinate points in the three-dimensional data generated by pulmonary nodules
were selected to generate the corresponding Mask map to provide training samples
for subsequent pulmonary nodules detection.

\section{methods}
\subsection{Pulmonary Nodule Detection
}
\begin{figure}[H]
	\centering
	\includegraphics[width=1\textwidth]{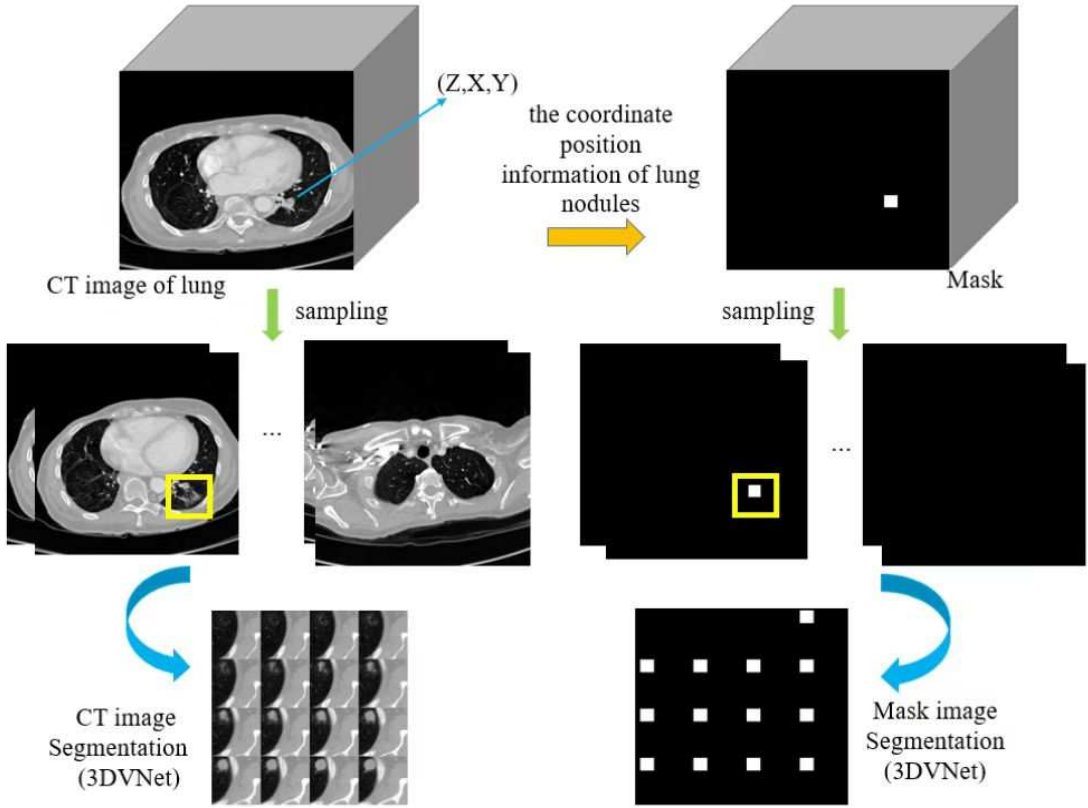}
	\caption{Detection algorithm flow.}
	\label{dect}
\end{figure}

The whole process of pulmonary nodules detection algorithm is divided into the following steps:
 firstly, a Mask for pulmonary nodules is generated according to the coordinates
 and diameters of nodules given by the Luna$16$ database;
 then, the CT image is denoised,
 CT scan (window width, 600 H; window level, $\textnormal{-}$680 H) 
 and the denoised image is normalized \cite{Radiology2002,Huang2019}.
 Then, we prepare the detection data of pulmonary nodules,
 and interpolated the CT images and the corresponding mask images.
 Then image segmentation was carried out to preserve the effective area of size
 including nodules and input it into 3DVNET \cite{milletari2016,yanglin2017} network to detect pulmonary nodules
 by training network.
 The detection process is shown in Figure~\ref{dect}.

\subsubsection{Accuracy Function
}

In this section, we introduce the calculation method of accuracy.
In the field of image segmentation, the Dice (coefficient function) is commonly used to measure the overlap of two samples\cite{Shamir2019}.Based on Dice function, we propose an accuracy calculation method. The gray value of the Mask image was used to determine whether the corresponding input image had nodules or not. 
If the input image has a Mask image, the predicted image also has a Mask image.
If the predicted Mask imag and the image of the real Mask image intersect, the prediction is correct,
otherwise it is judged as the prediction is wrong.
Let the number of accurate values be V\_right, and V\_all be the number of real samples with Mask. Accuracy function as shown in Formula (\ref{accf}).
\begin{equation}
	Acc=\frac{V\_right}{V\_all}
	\label{accf}
\end{equation}

\subsection{Classification of Pulmonary Nodule 
}
\begin{figure}[H]
	\centering
	\includegraphics[width=1\textwidth]{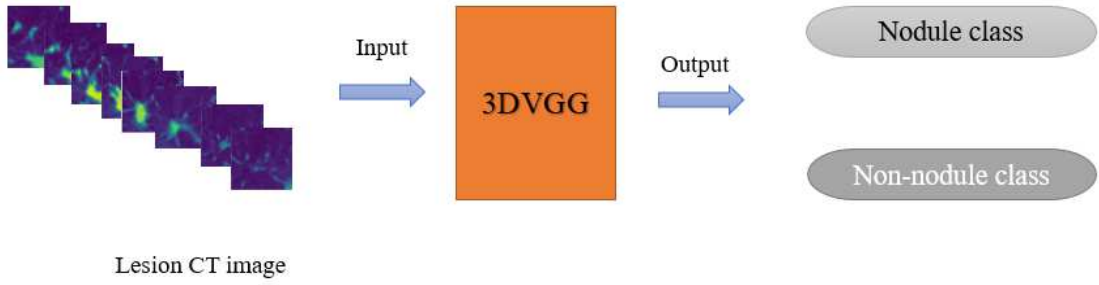}
	\caption{The classification flow chart.}
	\label{class}
\end{figure}

In this section, we introduce a classification model for pulmonary nodules \cite{simonyan2014}.
 Takes a sequence of pulmonary nodules image blocks as the input of 3DVGG network,
 and classifies the true and false nodules through 3DVGG network.
 The model training process of classification method is shown in Figure~\ref{class}.

\subsection{Pixels Embedding}
In this section, we introduce of pixel embedding. 
In image classification tasks,
 the network is sensitive to noise.
 In order to overcome this problem,
 an image with noise is added to the training database to train the network to distinguish the noise.
 Based on our previous work,
 we tested the MNIST database of PE hand-written numbers.
 By comparing the performance of the networks with PE and without PE on the test database of MNIST database,
 it is shown that the network with PE is better than the traditional network
 in the image classification task with noise,
 so it has higher noise resistance. 
 The robustness and training effect of the model were analyzed
 by introducing the embedding mechanism \cite{zhouchichun2020} in lung image detection and classification tasks.

\begin{figure}[H]
	\centering
	\includegraphics[width=1\textwidth]{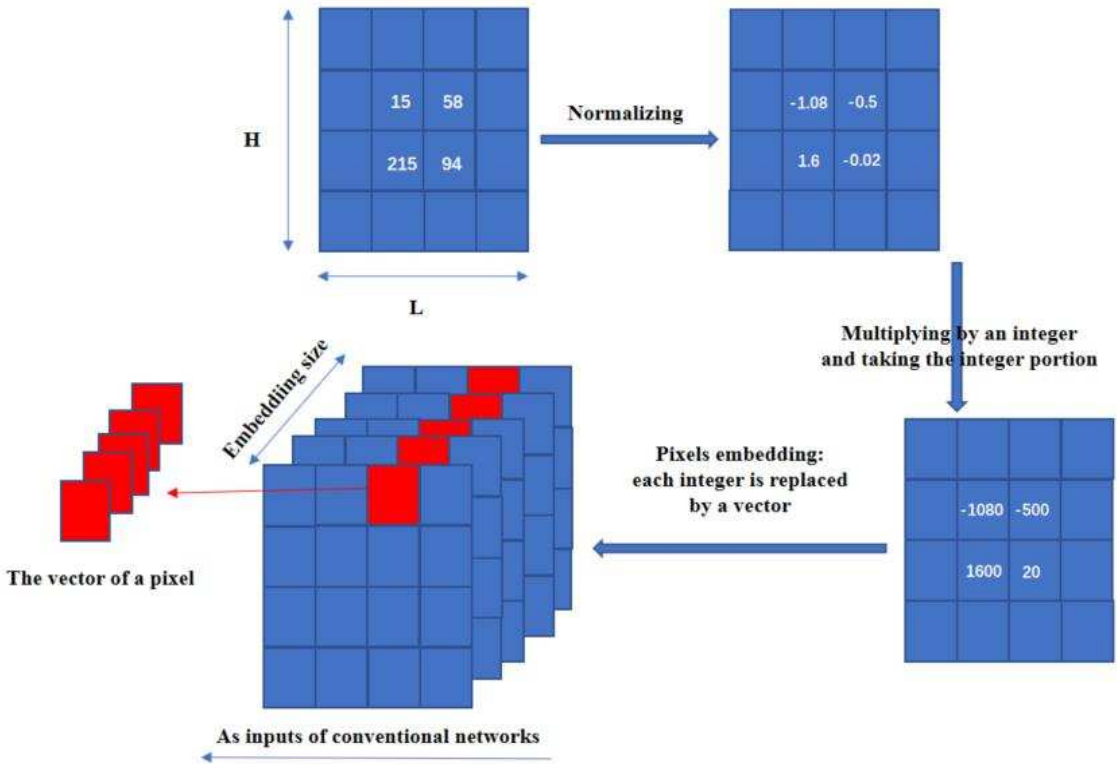}
	\caption{Pixel embedding steps.}
	\label{pe}
\end{figure}

The main technique of pixel embedding is to replace the single-valued pixel with a vector of shape $ 1\times M $ ,
 and the embedding size is $ M $ ,
 where the value of the vector is the parameter to be trained.
 Using pixel embedding mechanism,
 the network can "understand" the meaning of pixels and learn to distinguish noise automatically.The pixel embedding steps are shown in Figure~\ref{pe}.
 In this paper,
 the embedding mechanism was introduced into the network model of pulmonary nodules detection
 and recognition to observe the effect of the embedding mechanism on the training performance of the network.
 
\section{Results}

\subsection{Augmentation Level Experiments}
In this section, we explores the data augmentation method based on the generation model.
The datasets including the luna16 dataset and 
 the dataset generated by pulmonary nodule generation methods based on 3D pixel-level statistics and GAN framework.
In order to quickly get the results of the model evaluation by the data augmentation method,
 this paper selects the part of the data from LUNA16,
 including 2000 samples in one copy and 1000 samples in the other,
2000 samples were used as the base dataset.
 The specific Settings of data sources are shown in Table\ref{data31} and Table \ref{data_s}.

\begin{table}[H]
\centering
\caption{Data source: Dataset settings list.}
\centering
	\begin{tabular}{ll}     
\hline
$\makecell[c]{The\,Data\,Source}$&$\makecell[c]{Number\,of\,Training\,Samples}$\\
\hline
$\makecell[c]{Luna16}$& \makecell[c]{$2000+1000$}\\
\hline
$\makecell[c]{GAN\,generation\,dataset}$&\makecell[c]{$500$}\\
\hline
$\makecell[c]{Probability\,generates\,dataset}$&\makecell[c]{$500$}\\
\hline
\end{tabular}
\label{data_s}
\end{table}

\begin{figure}[H]
	\centering
	\includegraphics[width=1\textwidth]{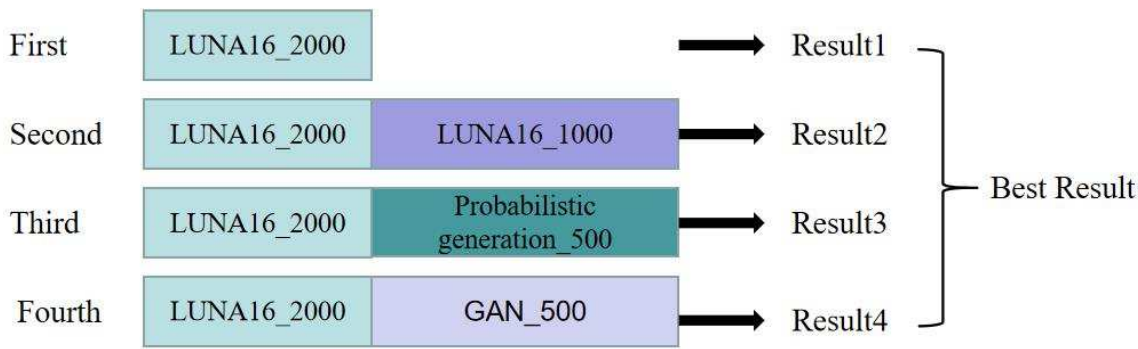}
	\caption{Classification verification.}
	\label{classver}
\end{figure}

Classified test experiments were performed for network optimization and overall evaluation.
2000 samples from Luna16 database were taken as basic control samples and named as original dataset 1,
 and 1000 samples from Luna16 were named as original dataset 2.
 The process of classifying test datasets is shown in Figure~\ref{classver}.

\begin{figure}[H]
	\centering
	\subfigure[Loss curve]{
		\includegraphics[width=0.48\textwidth]{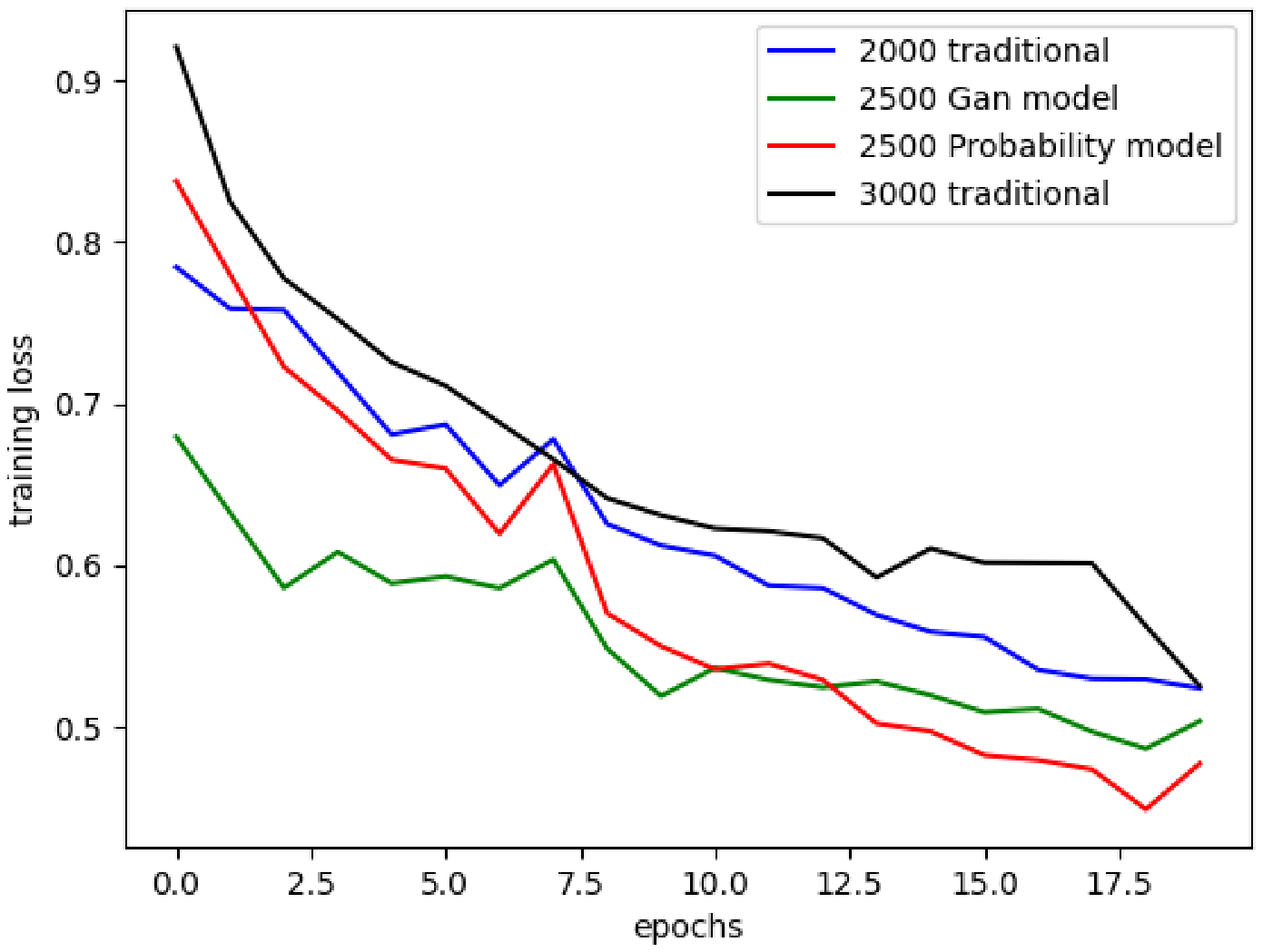}}
	\subfigure[Accuracy curve]{
		\includegraphics[width=0.48\textwidth]{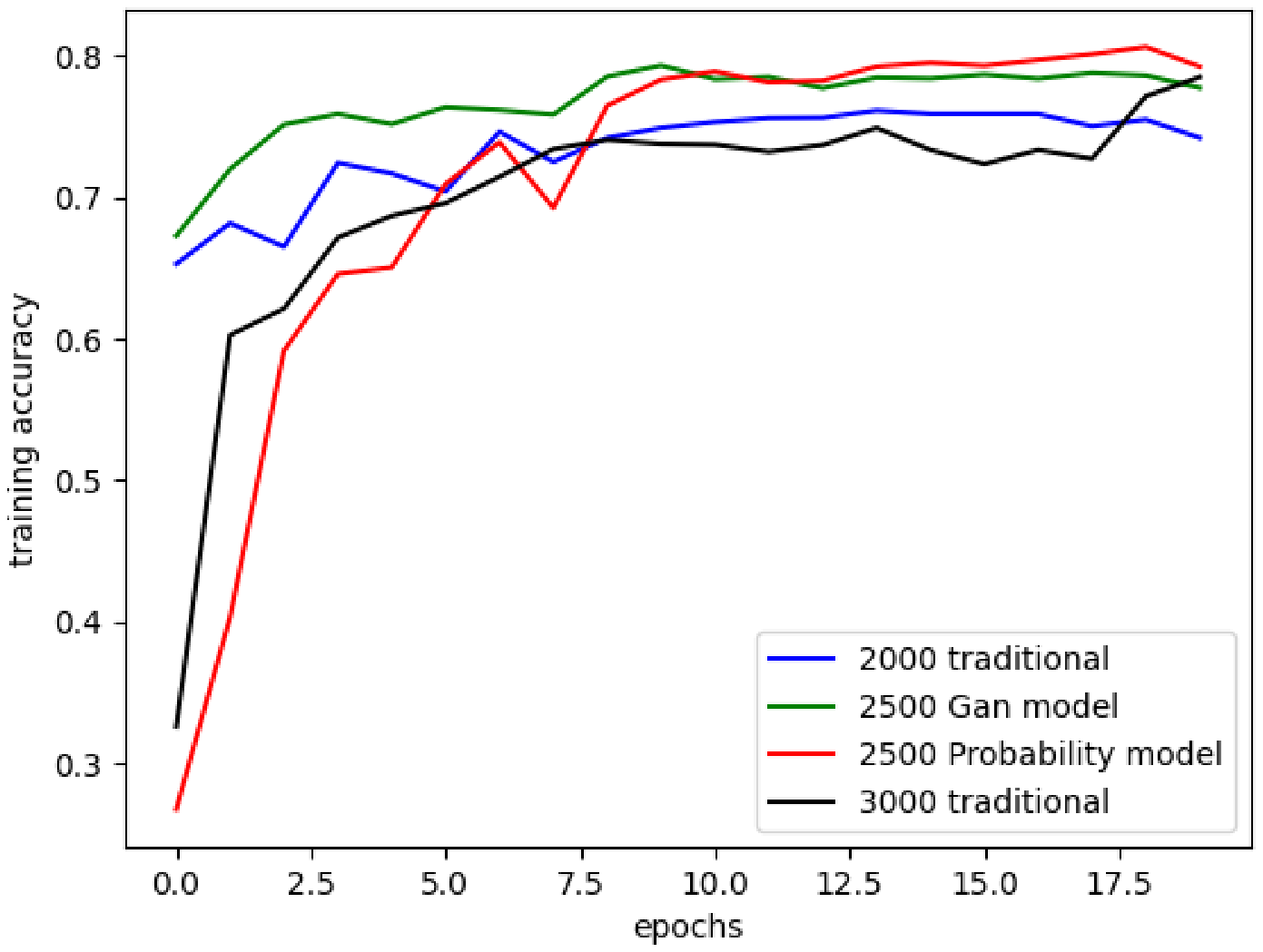}}
	\caption{Losses and accuracy curves.}
	\label{comparela}
\end{figure}

For better understanding of the training improvement results of the 3DVNET model
 based on the classification test dataset,
 a comparative analysis is demonstrated with both the accuracy of the training set and the forecast graph.
 The accuracy and loss function curves of 3DVNET models trained by different test sets
 are shown in Figure~\ref{comparela}.
 The relevant parameters in the figure:
2000 traditional represents dataset 1(LUNA16\_2000); 3000 traditional represents dataset2(LUNA16\_2000 and LUNA16\_1000); The 2500 GAN model represents the model dataset3(LUNA16\_2000 and GAN\_500).
2500 Probabilit Model represents the dataset 4(LUNA16\_2000 and probabilistic generation\_500). The accuracy and loss values of 3DVNET model trained by different data sets are shown in Table \ref{comla}.

\begin{table}[H]
	\centering
	\caption{The experimental results of different dataset detection tasks.}
	\centering
	\begin{tabular}{ccc}     
		\hline
		$\makecell[c]{The\,dataset}$&$\makecell[c]{Loss\,function value}$&$\makecell[c]{Accurate\,value}$\\
		\hline
		$\makecell[c]{Dataset1}$& \makecell[c]{$0.586$}& \makecell[c]{$0.751$}\\
		\hline
		$\makecell[c]{Dataset2}$&\makecell[c]{$0.525$}& \makecell[c]{\textcolor[rgb]{1,0,0}{$0.785(3.4\%\uparrow)$}}\\
		\hline
		$\makecell[c]{Dataset3}$&\makecell[c]{$0.503$}& \makecell[c]{\textcolor[rgb]{1,0,0}{$0.777(2.6\%\uparrow)$}}\\
		\hline
		$\makecell[c]{Dataset4}$&\makecell[c]{$0.477$}& \makecell[c]{\textcolor[rgb]{1,0,0}{$0.792(4.1\%\uparrow)$}}\\
		\hline
	\end{tabular}
	\label{comla}
\end{table}

As can be intuitively seen from Figure~\ref{comparela},
 both loss curve and accuracy curve show a stable trend.
As can be seen from Table \ref{comla},
 the accuracy rate of 0.751 was obtained by using the basic dataset 1.
 On the basis of dataset 1,
 and the accuracy rate of dataset 2 was increased by 3.4$\%$,
 proving that the increase of the amount of data is conducive to improving the accuracy rate of deep learning model.
In dataset 3, 500 samples from the GAN model were added to the basic dataset,
 and the accuracy was improved by 2.6\%.
In dataset 4,
 500 samples from the 3D pixel-level statistics algorithm were added to the basic dataset,
 and the accuracy was improved by 4.1\%.
 
In this section, we introduce the test results, the dataset are shown in the Table \ref{data32}. Under the same standard,
 the classification test results show that the generation algorithm
 has a great feasibility to improve the training effect of the pulmonary nodule model,
 and the pulmonary nodule generation algorithm based on 3D pixel-level statistics
 in this paper is the most effective in improving the training results of the network model.
As the whole experiment did not use the full dataset to train the model,
 the overall detection accuracy was not very high.
 The classification test results show that the generation algorithm
 has a great feasibility to improve the training effect of the pulmonary nodule model,
 and the pulmonary nodule generation algorithm based on 3D pixel-level statistics
 in this paper is the most effective in improving the training results of the network model.
 Compared with traditional data augmentation methods,
 the advantage of model generation method is that it can be widely used in other directions.

\subsection{Embedding Mechanism Level Experiments}
\begin{figure}[H]
	\centering
	\includegraphics[width=1\textwidth]{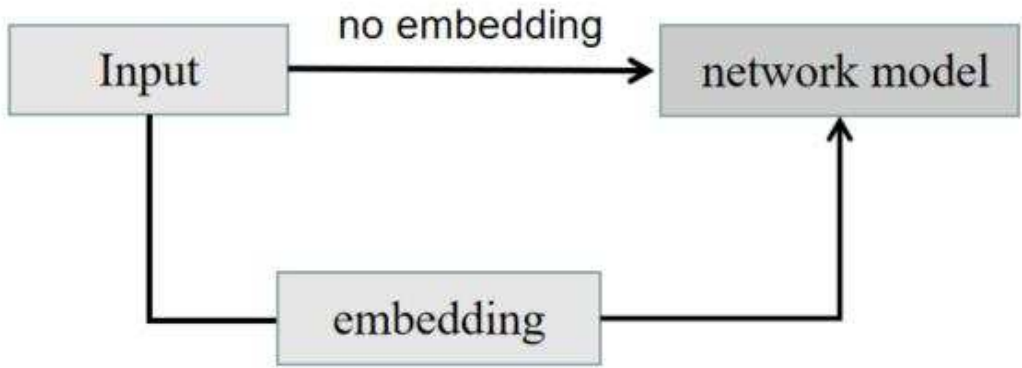}
	\caption{Model structure of implicit variable embedding mechanism.}
	\label{eim}
\end{figure}

Based on the previous work,
 we used the hidden variable embedding mechanism to test the MNIST database with handwritten numbers. 
 The noise-added MNIST test dataset is used to perform the classification task, 
 and the hidden variable embedding mechanism is introduced into the classification network.
 The results show that the network with the embedding mechanism
 is superior to the traditional network in the classification task of noisy images, 
 so the network with the embedding mechanism has better anti-noise performance. 
 Therefore, this paper applied pixel embedding technology to the detection
 and classification task of pulmonary nodules to analyze the robustness and training effect of the model. 
 We typically add a hidden variable embedding mechanism after the input layer, as shown in Figure~\ref{eim}.

\subsubsection{Add the Embedding Mechanism to the Detection Model}
Compared with the ordinary network,
 the network with the embedding mechanism is equivalent
 to expand the channel dimension characteristics of the image,
 and enhance the representability of the image.

\begin{figure}[H]
	\centering
	\subfigure[Loss curve]{
		\includegraphics[width=0.48\textwidth]{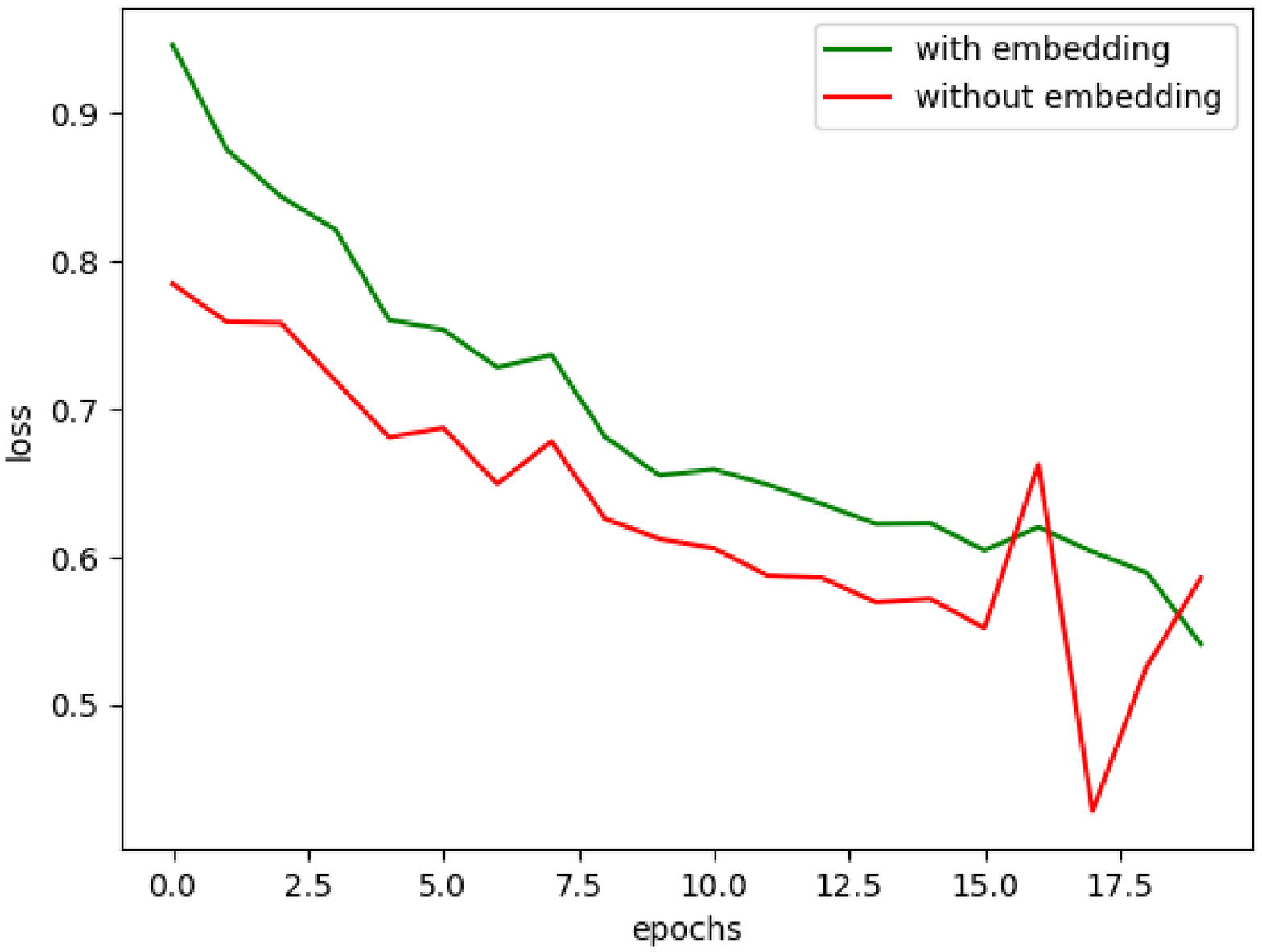}}
	\subfigure[Accuracy curve]{
		\includegraphics[width=0.48\textwidth]{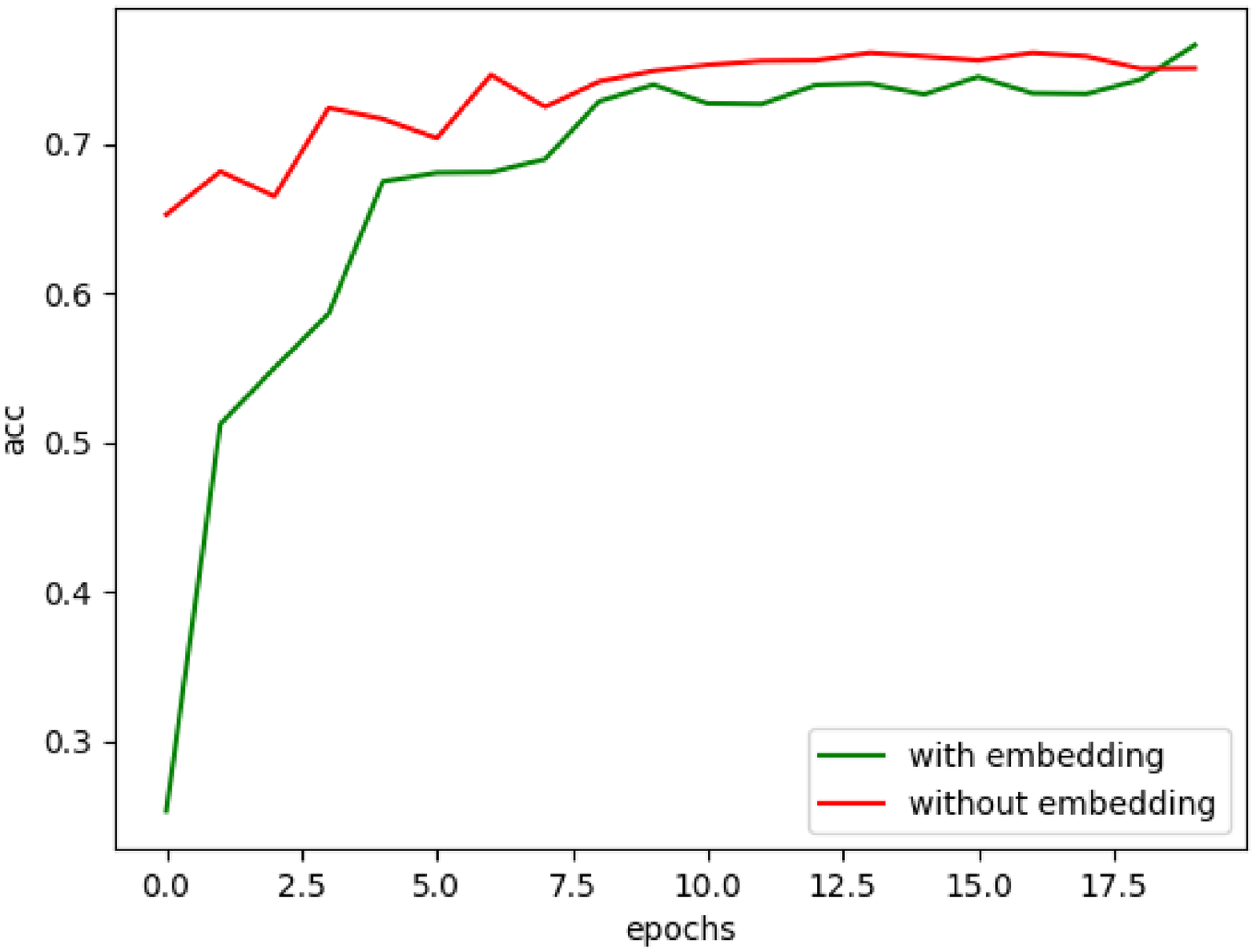}}
	\caption{Losses and accuracy curves.}
	\label{out19_1}
\end{figure}

In this section, we train the network with embedding mechanism.
The experimental dataset was trained to add the model of implicit variable embedding mechanism by dataset 1(LUNA16\_2000), and compared with the previous dataset 1 training results. The accuracy and loss function curves of the embedding mechanism model are shown in Figure~\ref{out19_1}.

\begin{table}[H]
	\centering
	\caption{The experimental results of different dataset detection tasks.}
	\centering
	\begin{tabular}{ccc}     
		\hline
		$\makecell[c]{Methods}$&$\makecell[c]{Loss\,function\, value}$&$\makecell[c]{Accurate\,value}$\\
		\hline
		$\makecell[c]{Without\, embedding}$& \makecell[c]{$0.586$}& \makecell[c]{$0.751$}\\
		\hline
		$\makecell[c]{With\,embedding }$&\makecell[c]{$0.541$}& \makecell[c]{\textcolor[rgb]{1,0,0}{$0.767(1.6\%\uparrow)$}}\\

		\hline
	\end{tabular}
	\label{ncomla}
\end{table}
As can be seen from Table \ref{ncomla},
 the use of embedding mechanism in 3DVNET network can improve the accuracy of the network,
 and the accuracy is relatively increased by $3.3\%$.
 From the perspective of experiment,
 this section proves that introducing the embedding mechanism into the network model
 of pulmonary nodules detection can improve the accuracy of model training.
 
 In this section, we test the network with embedding mechanism.
 The dataset are shown in the Table \ref{data32},
 the test dataset was derived from 4 cases in LUNA16 selected 649 positive samples and 7073 negative samples, 
50 positive samples and 50 negative samples are selected from the dataset for experiments.
The effect of the predicted Mask map obtained by the test results is poor, and the accuracy is low. 
It was concluded that there was overfitting in the training process, probably because not enough training data was used and only selected
Thirteen out of 888 cases in the LUNA16 dataset are used as experimental samples.
\subsubsection{Add the Embedding Mechanisms to the Classification Model}
The embedding mechanism can be further applied in other medical imaging tasks.
In our experiment, We classified true and false pulmonary nodules.
 The train dataset are shown in the Table \ref{data33}, 
 and the test dataset are shown in the Table \ref{data34}, 
 12,500 positive samples and 24,042 negative samples obtained by using traditional data enhancement methods
 (such as rotating, and flipping input images) were selected for the training model,
 and the testset is 100 positive and negative samples evenly divided.
 The effect of the embedding mechanism added to the training network
 on the performance improvement of the model was analyzed.
 
\begin{figure}[H]
	\centering
	\subfigure[Without embedding]{
		\includegraphics[width=0.48\textwidth]{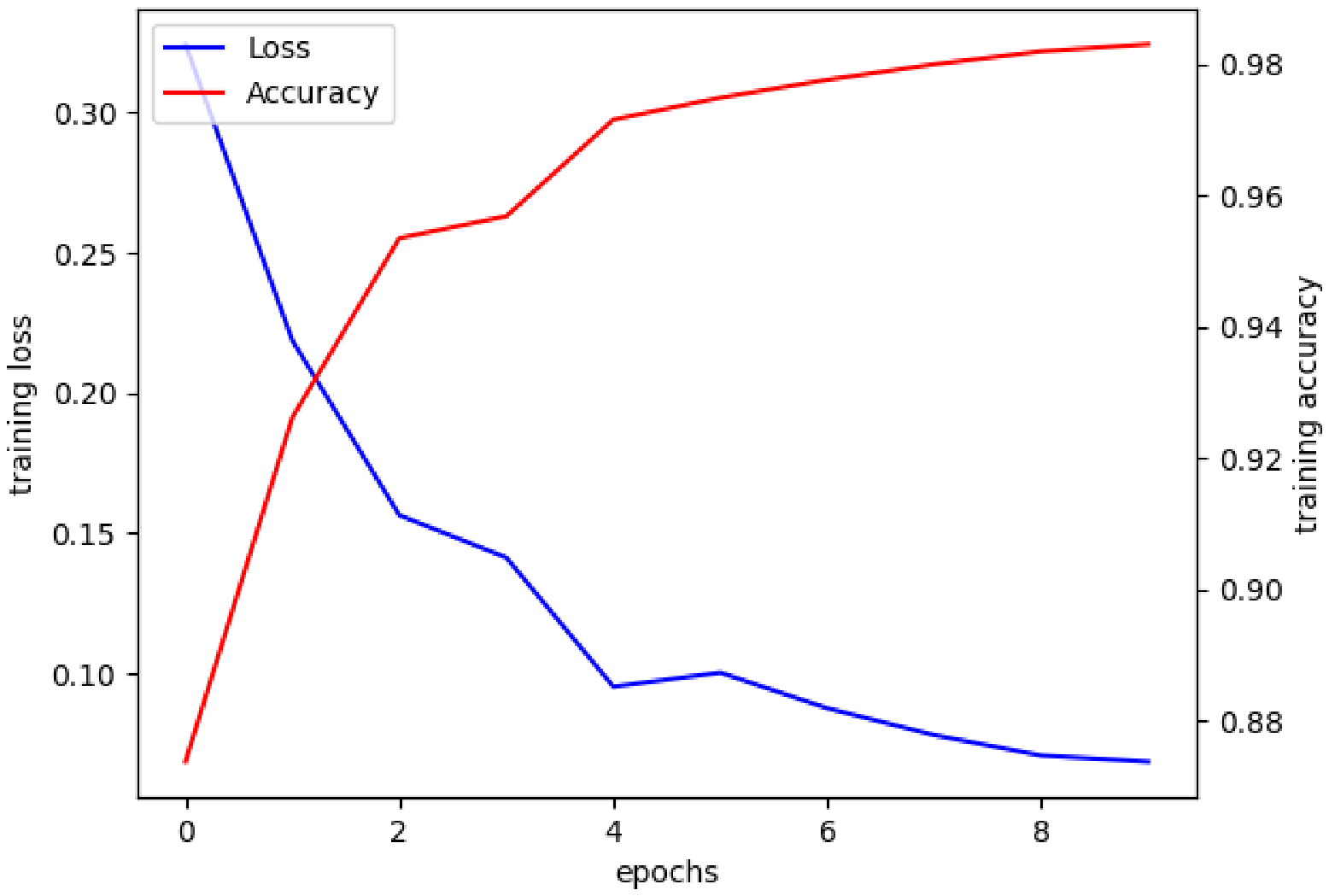}}
	\subfigure[[With embedding]{
		\includegraphics[width=0.48\textwidth]{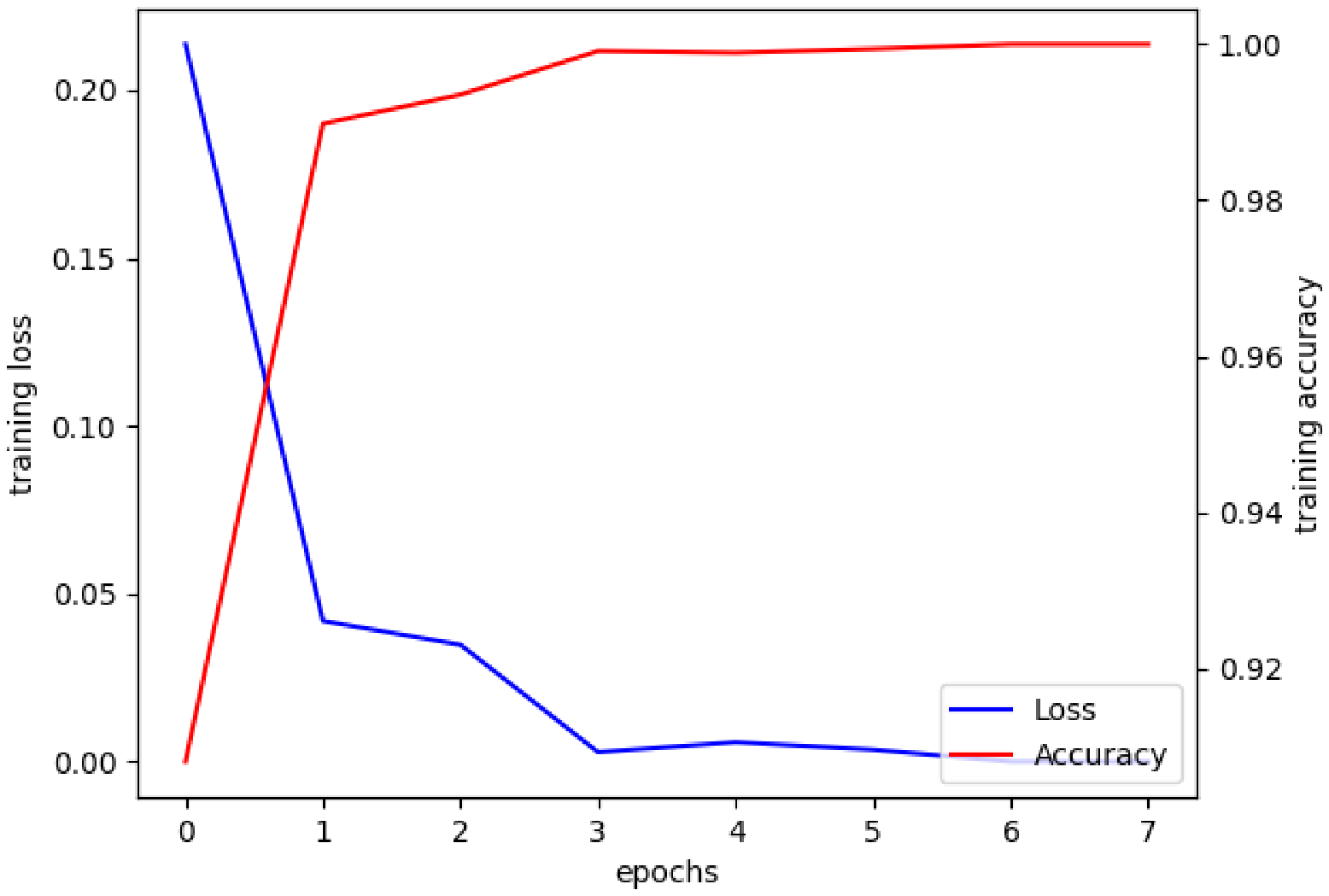}}
	\caption{Accuracy and loss curves of the training model.}
	\label{embeddingc}
\end{figure}

In this section, we introduce some evaluation metrics for binary classification.
The performance of binary classification can be quantitatively determined by $Precision$, $ Recall $, $ Accuracy $, and $ F1-score $. $ Accuracy $ rate refers to the proportion of positive examples in the examples classified as positive examples, and $ recall $ rate measures the proportion of positive examples classified as positive examples. $ F1 score $ is the harmonic average of accuracy rate and recall rate, which is often used as the final evaluation index in classification competition. They can be calculated using true positives $ (TP) $, true negatives $ (TN) $, false negatives $ (FN) $, and false positives $ (FP) $ , The calculation formula is shown below:	
	\begin{align}
		Precision &=\dfrac{TP}{TP+FP}\\
		Recall &=\dfrac{TP}{TP+FN}\\
		Accuracy &=\dfrac{TP+TN}{TP+TN+FP+FN}\\
		F1-score &=\dfrac{2TP}{2TP+FP+FN}	
	\end{align}
The improvement effect of model performance with and without embedding mechanism
in the training network was compared and analyzed.
The accuracy and loss function curves of the training process without embedding
and with embedding are shown in Figure~\ref{embeddingc}.

\begin{figure}[H]
	\centering
	\subfigure[Without embedding]{
		\includegraphics[width=0.48\textwidth]{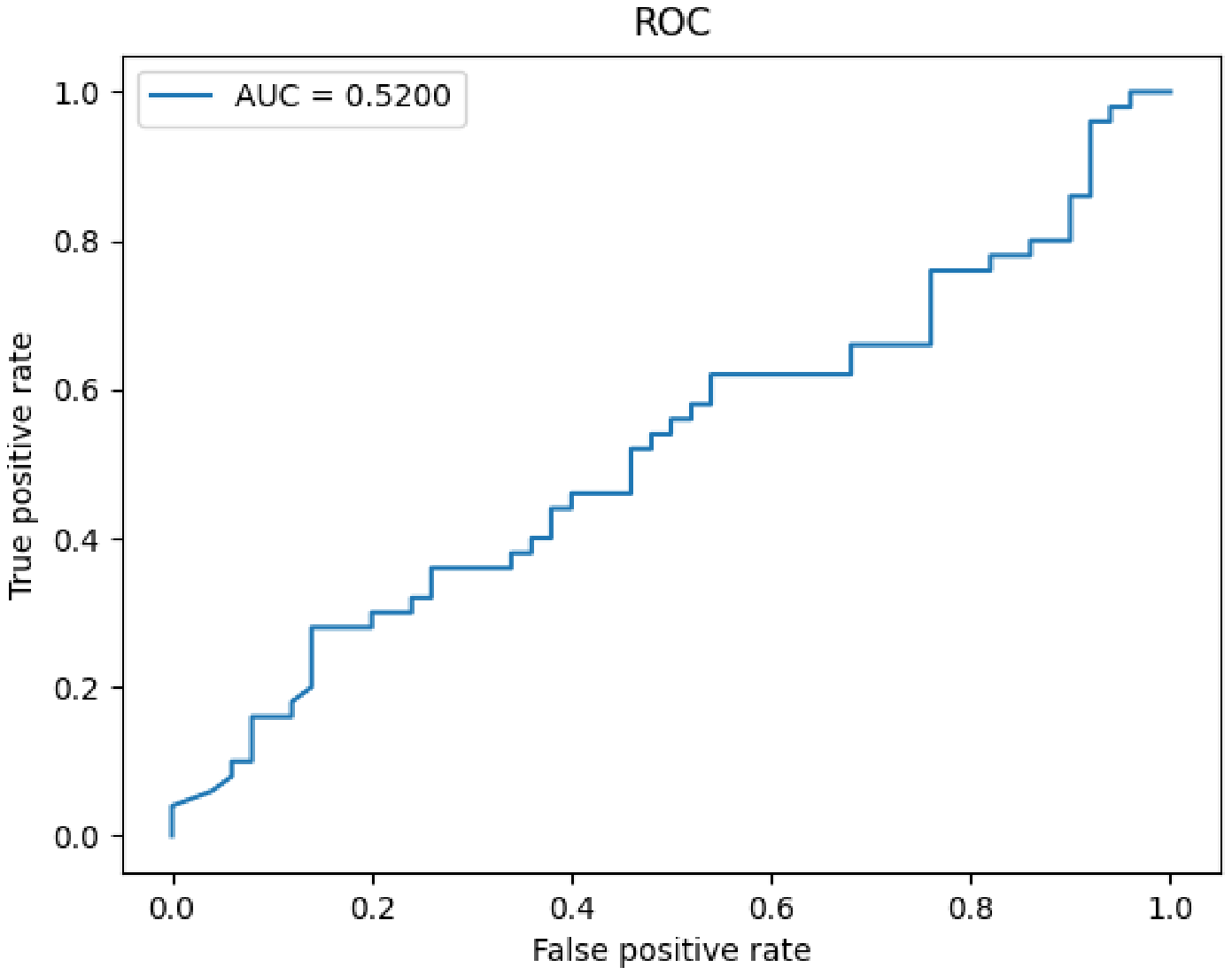}}
	\subfigure[With embedding]{
		\includegraphics[width=0.48\textwidth]{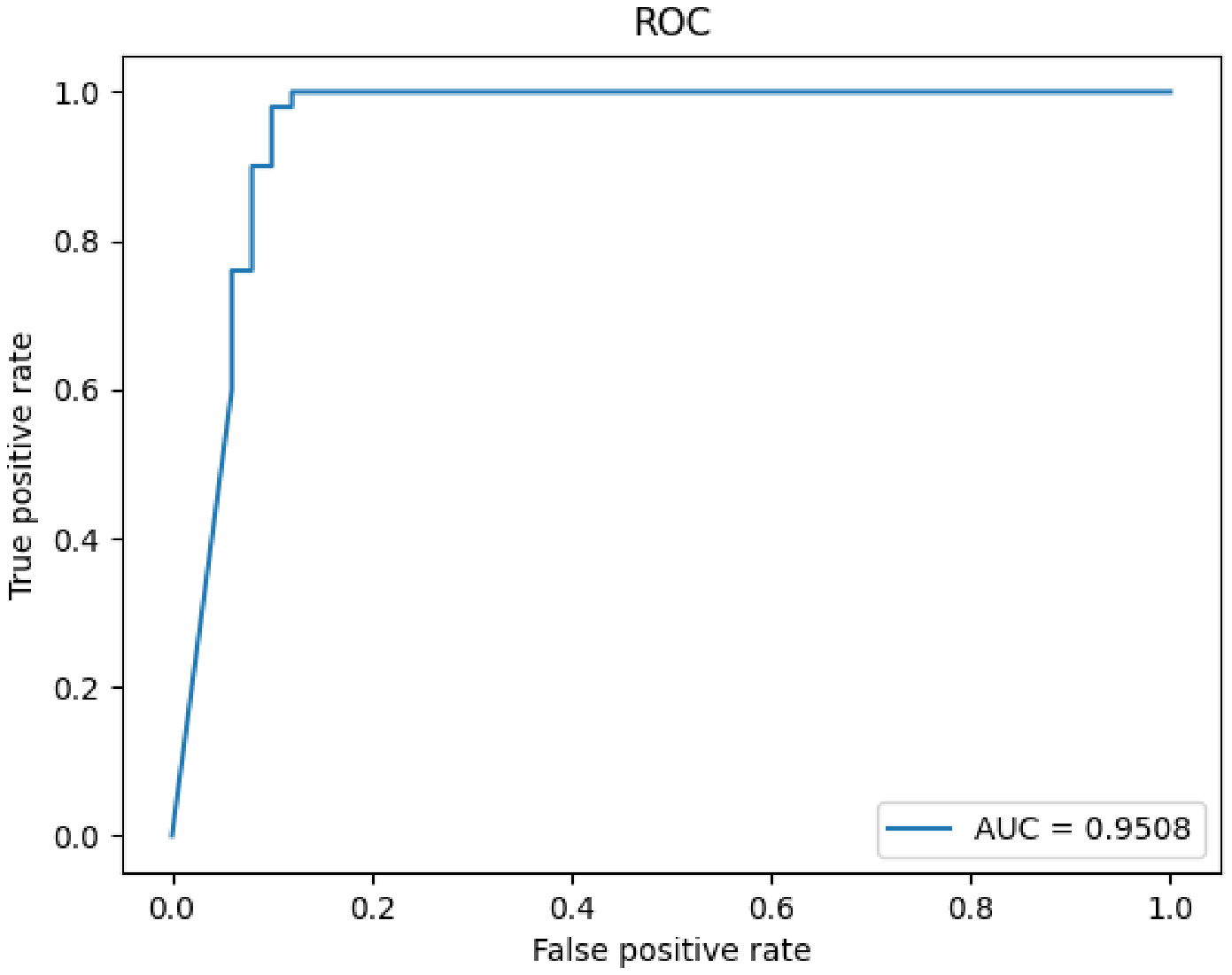}}
	\caption{The ROC curve.}
	\label{roc}
\end{figure}

As can be seen from Figure~\ref{embeddingc},
 with the increase in training times,
 the accuracy of both the training process without Embedding and Embedding increased with high values.
 The loss curve and the accuracy curve showed a steady trend,
 and the accuracy of model training with Embedding is close to 1.
The experimental results of without embedding and with embedding training process are shown
 in Table \ref{confusion} to
 Table \ref{eindex_2} and Figure~\ref{roc}.

\begin{table}[H]
	\centering
	\caption{Confusion matrix.}
	\centering
	\begin{tabular}{cccc}     
		\hline
		$\makecell[c]{Methods}$&  \diagbox{$Truth$}{$Prediction$}  &$\makecell[c]{Positive}$ &$ \makecell[c]{ Negative} $\\
		\hline
		\multirow{2}*{$Without\,embedding$} & $ \makecell[c]{Positive} $ & \makecell[c]{$42(TP)$} & $ \makecell[c]{8(FN)} $ \\
		~ & $ \makecell[c]{Negative} $ & $ \makecell[c]{46(FP) }$ & $ \makecell[c]{4(TN) }$ \\
		\hline
		\multirow{2}*{$With\,embedding$} & $ \makecell[c]{Positive} $ & \makecell[c]{$50(TP)$} & $ \makecell[c]{0(FN)} $ \\
		~ & $ \makecell[c]{Negative} $ & $ \makecell[c]{10(FP) }$ & $ \makecell[c]{40(TN) }$ \\
		\hline
	\end{tabular}
	\label{confusion}
\end{table}

\begin{table}[H]
	\centering
	\caption{The evaluation index of true pulmonary nodules.}
	\centering
	\begin{tabular}{ccccc}     
		\hline
		$\makecell[c]{Methods}$&  $\makecell[c]{Precision}$ &$\makecell[c]{Recall}$ &$ \makecell[c]{ Accuracy} $ & \makecell[c]{F1-score } \\
		\hline
		$\makecell[c]{Without\\embedding}$&  $\makecell[c]{0.48}$ &$\makecell[c]{0.84}$ &$ \makecell[c]{0.46} $ & \makecell[c]{0.609 } \\
		\hline
		$\makecell[c]{With\,embedding}$&  \makecell[c]{\textcolor[rgb]{1,0,0}{$0.83(35.0\% \uparrow)$}}  & \makecell[c]{\textcolor[rgb]{1,0,0}{$1.00(16.0\% \uparrow)$}} & \makecell[c]{\textcolor[rgb]{1,0,0}{$0.90(44.0\% \uparrow)$}}&\makecell[c]{\textcolor[rgb]{1,0,0}{$0.91(30.0\% \uparrow)$} } \\
		\hline
	\end{tabular}
	\label{eindex}
\end{table}

\begin{table}[H]
	\centering
	\caption{The evaluation index of false pulmonary nodules.}
	\centering
	\begin{tabular}{ccccc}     
		\hline
		$\makecell[c]{Methods}$&  $\makecell[c]{Precision}$ &$\makecell[c]{Recall}$ &$ \makecell[c]{ Accuracy} $ & \makecell[c]{F1-score } \\
		\hline
		$\makecell[c]{Without\\embedding}$&  $\makecell[c]{0.33}$ &$\makecell[c]{0.08}$ &$ \makecell[c]{0.46} $ & \makecell[c]{0.13} \\
		\hline
		$\makecell[c]{With\,embedding}$&  \makecell[c]{\textcolor[rgb]{1,0,0}{$1.00(67.0\% \uparrow)$}}  & \makecell[c]{\textcolor[rgb]{1,0,0}{$0.80(72.0\% \uparrow)$}} & \makecell[c]{\textcolor[rgb]{1,0,0}{$0.90(44.0\% \uparrow)$}}&\makecell[c]{\textcolor[rgb]{1,0,0}{$0.89(53.0\% \uparrow)$} } \\
		\hline
	\end{tabular}
	\label{eindex_1}
\end{table}

\begin{table}[H]
	\centering
	\caption{Average evaluation index.}
	\centering
	\begin{tabular}{ccccc}     
		\hline
		$\makecell[c]{Methods}$&  $\makecell[c]{Precision}$ &$\makecell[c]{Recall}$ &$ \makecell[c]{ Accuracy} $ & \makecell[c]{F1-score } \\
		\hline
		$\makecell[c]{Without\\embedding}$&  $\makecell[c]{0.41}$ &$\makecell[c]{0.46}$ &$ \makecell[c]{0.46} $ & \makecell[c]{0.37 } \\
		\hline
		$\makecell[c]{With\,embedding}$&  \makecell[c]{\textcolor[rgb]{1,0,0}{$0.92(51.0\% \uparrow)$}}  & \makecell[c]{\textcolor[rgb]{1,0,0}{$0.90(44.0\% \uparrow)$}} & \makecell[c]{\textcolor[rgb]{1,0,0}{$0.90(44.0\% \uparrow)$}}&\makecell[c]{\textcolor[rgb]{1,0,0}{$0.90(53.0\% \uparrow)$} } \\
		\hline
	\end{tabular}
	\label{eindex_2}
\end{table}

The whole test dataset contains 100 samples, which are divided into 50 false nodules and 50 true nodules. 
The results show that the average recall rate is 0.46 and the average accuracy rateis 0.41 without embedding model test dataset.
The average recall rate with embedding is 0.90,
the average accuracy rateis 0.92, and the average F1-score is 0.90.
With embedding in the process of training mechanism model can obviously improve training effect, 
the average accuracy rate  increased 44.0$\%$,
the average precision rate increased by 51.0$\%$, 
and the training time also almost 50$\%$ increased.

\section{Discussion}

Lung cancer screening by CT scan is the most effective way to detect early cancer.
Because of the small amount of data in medical field and the collection of medical data is limited by ethical privacy, We need to  start with  with small samples studies.
 Suspected pulmonary nodule areas were extracted mainly using 3DVNET network,
 pulmonary nodules and non-nodules were classified using 3DVGG network.
 The accuracy and robustness of deep learning models for detection and classification of pulmonary nodules
 can be improved by increasing the data and improving the model structure in small samples.
 The main work of this paper is as follows:
\begin{itemize}
	\item [1.]
	The main aim of the present paper is two folds: 
	(1) proposing a new data augmentation 
	method and (2) introducing the embedding mechanism to improve the existing models.
	\item [2.]
	In the augmentation method, 
	a 3D pixel-level statistics algorithm 
	is proposed to generate pulmonary nodule samples. 
	The result of the 3DVNET model with the augmentation method for pulmonary nodule detection 
	shows that the proposed data augmentation method outperforms
	the method based on generative adversarial network (GAN) framework.
	\item [3.]
	The embedding mechanism are designed to better understand the meaning of pixels of 
	the pulmonary nodule samples by introducing hidden variables. 
	The result of 3DVGG network with embedding mechanism for pulmonary nodule classification 
	shows that the embedding mechanism improves the accuracy and robustness for 
	the classification of pulmonary nodules obviously. 
	The $accuracy$ of the model testset reaches 84$\%$, 
	and the $recall$ rate reaches 1, 
	the high score of $F1-score$ was 0.909. 
	The proposed data augmentation method and embedding mechanism
	can be further applied in other medical imaging tasks.

\end{itemize}

According to the research,
small samples may be enough to show the presence of an effect but not for estimating the effect size. 
If the objective is only to show that an effect exists, bearing the cost of a large sample can be avoided.
Because of the small amount of data in medical field, it is of practical significance to use small samples for research.
 
 In conclusion, we improved the deep learning algorithm with data augmentation method and embedding mechanism for pulmonary nodules classification on small samples, it would be utilized by physicians to enhance their performance on genuine nodules detection. Moreover, the improvement of deep learning can be further introduced to other image based diagnosis.












\end{document}